\documentclass[twocolumn,showpacs,preprintnumbers,amsmath,amssymb]{revtex4}

\usepackage{graphics}
\usepackage{dcolumn}
\usepackage{bm}
\usepackage{braket}

\begin{document}

\preprint{******}

\title{Entanglement of Orbital Angular Momentum States\\
between an Ensemble of Cold Atoms and a Photon}

\author{R. Inoue$^{1}$, N. Kanai$^{1}$, T. Yonehara$^{1}$, Y. Miyamoto$^{2}$, M. Koashi$^{3}$, and M. Kozuma$^{1,4}$}
\affiliation{%
$^{1}$Department of Physics, Tokyo Institute of Technology, 2-12-1 O-okayama, Meguro-ku, Tokyo 152-8550, Japan}
\affiliation{%
$^{2}$Department of Information and Communication Engineering, The University of Electro-Communications,\\
1-5-1 Chofugaoka, Chofu, Tokyo 182-8585, Japan}
\affiliation{%
$^{3}$Division of Materials Physics, Graduate School of Engineering Science, Osaka University,\\
1-3 Machikaneyama, Toyonaka, Osaka 560-8531, Japan}
\affiliation{%
$^{4}$PRESTO, CREST, Japan Science and Technology Agency, 1-9-9 Yaesu, Chuo-ku, Tokyo 103-0028, Japan}%

\date{\today}

\begin{abstract}
Recently, atomic ensemble and single photons were successfully entangled by using collective enhancement [D. N. Matsukevich, \textit{et al.}, Phys. Rev. Lett. \textbf{95}, 040405(2005).], where atomic internal states and photonic polarization states were correlated in nonlocal manner. Here we experimentally clarified that in an ensemble of atoms and a photon system, there also exists an entanglement concerned with spatial degrees of freedom. Generation of higher-dimensional entanglement between remote atomic ensemble and an application to condensed matter physics are also discussed.
\end{abstract}

\pacs{03.65.Wj, 03.67.Mn, 32.80.-t, 42.50.Dv}
\maketitle
\section{INTRODUCTION}
Higher-dimensional bipartite entangled states enable us to achieve more efficient quantum information processing \cite{Bjork2001,Peres2000}, for example by enhancing optical data traffic in quantum communications compared to usually employed two-dimensional entanglement between qubits.  Such higher-dimensional entangled states can be created by using Laguerre-Gaussian (LG) modes, since photons in LG modes carry orbital angular momenta (OAM) which can be utilized to define an infinite-dimensional Hilbert space \cite{Bagan2006}.  Inspired by the pioneering experiment using parametric down-conversion (PDC) \cite{AZ2001}, various protocols have been demonstrated using the OAM states of photons \cite{AZ2003-2,White2004}.
  
Recently, two-dimensional entanglement of a collective atomic excitation and a single photon has been successfully generated \cite{Kuzmich2005-1} using `DLCZ' scheme \cite{DLCZ2001}.  Two-dimensional entanglement of two remote atomic ensembles has also been reported \cite{Kimble2005,Kuzmich2006}.  However, so far there has been no discussion about important aspect related to spatial degrees of freedom of a collective atomic excitation and a photon.  In this paper, we report entanglement of OAM states between these two carriers of information.
While the measurement basis utilized in this study is limited to two dimensions, it is natural to presume that our system possesses higher-dimensional entanglement over a wide range of OAM.  Producing higher-dimensional entanglement between two remote long-lived massive atoms is an important step toward the realization of intricate tasks of quantum information.

\section{THEORETICAL MODEL}
\begin{figure}[h]
  \begin{center}
    \includegraphics{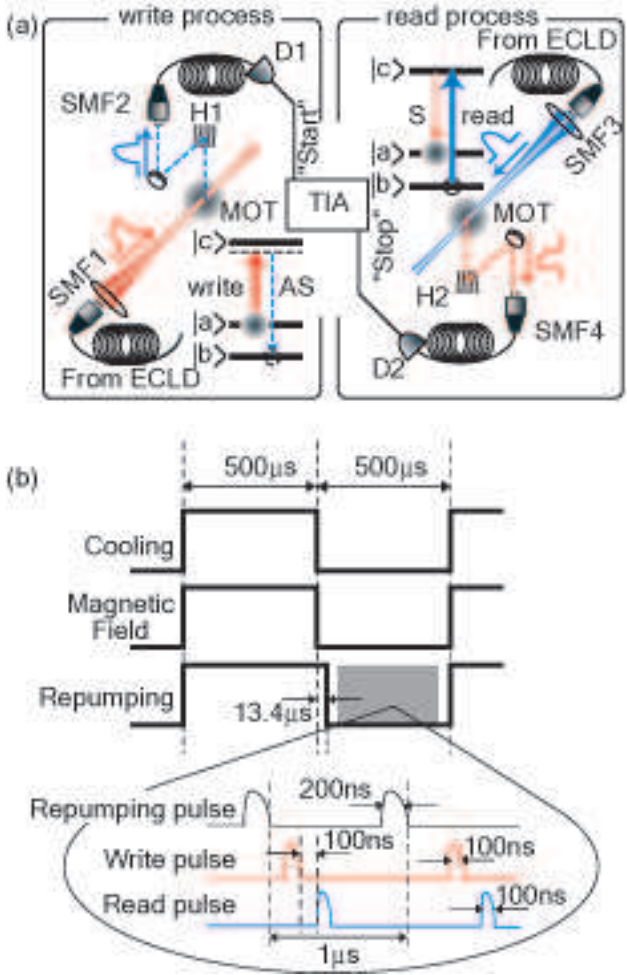}
  \end{center}
  \caption{(color online).  (a) Schematic of experimental setup.  \textit{Write} and \textit{read} pulses propagate into a cloud of cold $^{87}\mathrm{Rb}$ atoms (MOT), and generate the correlated output pair of anti-Stokes (AS) and Stokes (S) photons.  SMF1, 2, 3, and 4, single mode fibers;  H1 and H2, computer generated holograms;  D1 and D2, detectors;  TIA, time interval analyzer.  (b) Depicts the timing sequence for data acquisition.} 
  \label{fig:1.eps}
\end{figure}

Figure 1-(a) shows a schematic of our experimental setup.  We consider ensemble of atoms having the level structure of $\ket{a},\  \ket{b}\  \mathrm{and}\  \ket{c}$. Initially, all of the atoms are prepared in level $\ket{a}$. A classical \textit{write} pulse tuned to the $\ket{a}\rightarrow\ket{c}$ transition with proper detuning $\Delta$ is emitted from  a single mode fiber (SMF1) and is incident on the atomic ensemble.  In this process, the $\ket{c}\rightarrow\ket{b}$ transition is induced and an anti-Stokes photon is generated.  The \textit{write} pulse is weak and its interaction time is short so that the probability of scattering one anti-Stokes photon into a specified mode is much less than unity for each pulse.  We are interested in LG ($\mathrm{LG}_\mathit{pm}$) modes and a Gaussian ($\mathrm{LG}_\mathit{00}$) mode.  Linearly polarized light having well-defined orbital angular momentum can be described by $\mathrm{LG}_\mathit{pm}$ modes with the two indices, $p$ and $m$, $\textit{p}+1$ is the number of radial nodes and the azimuthal mode index $m$ identifies the number of the $2\pi$-phase shifts along a closed path around the beam center.  In our experiment, we consider only $\mathrm{LG}_\mathit{0m}$ modes, i.e. the radial index $p=0$.  $\mathrm{LG}_\mathit{0m}$ modes have a doughnut-shape intensity distribution and carry corresponding OAM of $m\hbar$ per photon \cite{Bagan2006}, which can be utilized to constitute an infinite-dimensional basis.  Mode detection of the anti-Stokes photon is performed using a computer generated hologram (H1) and a single-mode fiber (SMF2).  We can also analyze superpositions of an $\mathrm{LG}_\mathit{00}$ mode and an $\mathrm{LG}_\mathit{01}$ mode by shifting the dislocation of the hologram out of the center of the beam by a certain amount \cite{AZ2002}.  Here we define $\ket{m}_\mathrm{AS}$ to be the state of an anti-Stokes photon with OAM of $m\hbar$.

In the ideal case, detection of a single photon in a certain spatial mode results in the ensemble of atoms containing exactly one excitation in the corresponding collective atomic mode.  The collective atomic spin excitation operator is successfully defined as $\hat{S}^\dagger (\varphi)$, where  $\varphi(\bm{\mathrm{r}})$ corresponds to the spatial distribution of the relative phase between the \textit{write} pulse and the detected anti-Stokes photon.

The excited state of the atomic ensemble is given by $\hat{S}^\dagger (\varphi)\ket{i}_{a}$, where the initial state of the ensemble is $\ket{i}_{a}\equiv\otimes_{j}\ket{a_{j}}$.  The atomic ensemble therefore memorizes the relative phase between the \textit{write} pulse and the anti-Stokes photon.  The \textit{write} pulse has the mode profile of $\mathrm{LG}_\mathit{00}$ and propagates in the direction of wave vector $\bm{\mathrm{k}}_\mathrm{W}$, which is determined by SMF1.  Here we only consider the anti-Stokes photon propagating in the direction of $\bm{\mathrm{k}}_\mathrm{AS}$ determined by SMF2, which is almost the same as the direction of $\bm{\mathrm{k}}_\mathrm{W}$.  Then, the relative phase $\varphi(\bm{\mathrm{r}})$ is determined by the spatial mode profile of the detected anti-Stokes photon. 

Suppose that the detected photon has the mode of $\mathrm{LG}_\mathit{0m}$, 
and let $\varphi_{m}(\bm{\mathrm{r}})$ be the relative phase in this case.
The atomic ensemble is then in the excited state $\ket{-m}_{a}\equiv\hat{S}^\dagger (\varphi_{m})\ket{i}_{a}$, which has the angular momentum with the opposite sign, $-m\hbar$,  due to the angular momentum conservation law.  The variation of $\varphi_{m}(\bm{\mathrm{r}})$ along a closed path around the anti-Stokes beam axis is $2m\pi$, implying that the atomic ensemble memorizes the phase structure of the emitted anti-Stokes photon including singularity. 
Considering angular momentum conservation law between anti-Stokes photon and an atomic ensemble, the whole state of these is a coherent superposition of angular momentum states, which is written as 
\begin{eqnarray}
\ket{\Psi}_\mathrm{AS\& a}&=&\ket{\mathrm{vac}}_\mathrm{AS}\ket{i}_{a}\\
&+&\sqrt{p}\!\!\!\sum_{m=-\infty}^{\infty}\!\!\!\!C_{m}\hat{a}_{m}^\dagger\hat{S}^\dagger (\varphi_{m})\ket{\mathrm{vac}}_\mathrm{AS}\ket{i}_{a}+\mathrm{O}(p),\nonumber
\end{eqnarray}
where $p$ and $\ket{\mathrm{vac}}_\mathrm{AS}$ represent the excitation probability and the vacuum state of the anti-Stokes photon.  
The amplitude $C_{m}$ for each OAM state is expected to depend on the spatial shape of the region where the \textit{write} pulse interacts with the atoms \cite{Torner2003}.  In the case of negligibly small $\mathrm{O}(p)$, we obtain
\begin{eqnarray}\label{state}
\ket{\Psi}_\mathrm{AS\& a}&=&\ket{\mathrm{vac}}_\mathrm{AS}\ket{i}_{a}\nonumber\\
&+&\sqrt{p}\!\!\!\sum_{m=-\infty}^{\infty}\!\!\!\!C_{m}\ket{m}_\mathrm{AS}\ket{-m}_{a}.
\end{eqnarray}
This is a probabilistic higher-dimensional entangled state between an atomic ensemble and a photon.
In the present work, we investigated the entanglement concerned with $m=0$ and $m=1$ and thus experimentaly expected entangled state can be written down as
\begin{eqnarray}\label{expected}
\ket{\Psi(\gamma)}=\frac{1}{\sqrt{1+|\gamma|^2}}\biggl(\ket{0}_\mathrm{AS}\ket{0}_{a}+\gamma\ket{+1}_\mathrm{AS}\ket{-1}_{a}\biggl),
\end{eqnarray}
where complex number $\gamma$ represents relative phase and amplitude.  

The entanglement shown in Eq.(\ref{expected}) can be verified as follows. 
After a variable delay $\delta t$, we convert the atomic excitation to a single photon (Stokes photon) by illuminating the atomic ensemble with a laser pulse (\textit{read} pulse) resonant on the $\ket{b}\rightarrow\ket{c}$ transition.  Note that the efficiency of this transfer can be very close to unity because it corresponds to the retrieval process based on electromagnetically induced transparency \cite{Lukin2002}.  Atom-photon entanglement, therefore, can be checked by measuring quantum correlation between the anti-Stokes and Stokes photons.  
Since we detect the Stokes photon counter-propagating with respect to the anti-Stokes photon, the value of the OAM of the Stokes photon in the direction of propagation is equal to that of the anti-Stokes photon.  
We use $\ket{m}_\mathrm{S}$ to denote the state of a Stokes photon with OAM of $m\hbar$.  

\section{EXPERIMENT}
In the experiment, we created an optically thick (optical depth $\sim 4$) cold atomic cloud using a magneto-optical trap (MOT) of $^{87}\mathrm{Rb}$.  The levels $\{ \ket{a},\ket{b}\}$ correspond to $5S_{1/2}$, $F=\{ 2, 1\}$ levels, respectively, and the level $\ket{c} $ corresponds to $5P_{1/2}$, $F'=2$.  
The experimental sequence is shown in Fig.1-(b) schematically.  One cycle of our experiment comprised of a cooling period and a measurement period, and both periods had durations of $500\,\mathrm{\mu s}$.  After the cooling period, the magnetic field, the cooling and the repumping lights were turned off.  The repumping light was left on for an additional $13.4\,\mathrm{\mu s}$ after the cooling light was turned off, in order to prepare the initial state $\ket{i} $.  Our vacuum cell was magnetically shielded by a three-fold permalloy and the residual magnetic field was about $100\,\mathrm{\mu G}$.  However, influence of eddy current prolonged the decay time of the magnetic field, meaning that the residual magnetic field at MOT point during measurement process was actually about $200\,\mathrm{mG}$.
In the measurement period, the vertically polarized \textit{write} and \textit{read} pulses illuminated the atomic ensemble with a repetition cycle of $1\,\mathrm{\mu s}$.  Each cycle finished with illumination of $200\,\mathrm{ns}$ long repumping pulse to initialize the atomic ensemble.  The \textit{write} and \textit{read} pulses were produced by external cavity laser diodes (ECLDs), and their spatial modes were purified by passing through SMF1 and SMF3, respectively.  A $100\,\mathrm{ns}$ long \textit{write} pulse having $250\,\mathrm{nW}$ peak intensity was focused into the atomic ensemble with a Gaussian beam waist of $400\,\mathrm{\mu m}$, where the detuning of the \textit{write} pulse was set to $-10\,\mathrm{MHz}$.  Horizontally polarized anti-Stokes photons were coupled to a single mode fiber (SMF2) after being diffracted by the computer generated hologram (H1).  The coupling efficiency to SMF2 was maximized for the $\mathrm{LG}_\mathit{00}$ mode having a beam waist of $140\,\mathrm{\mu m}$ at the center of the atomic ensemble.  We utilized a reflection-type phase hologram blazed to increase the diffraction efficiency up to ${45}\,\mathrm{\%}$.  Two types of holograms were used in the experiment, one having no dislocation and the other having a single dislocation located at its center.  Our experiment using classical light demonstrated that the combination of a hologram and a single mode fiber could achieve mode detection with a distinction ratio of approximately $1000 : 1$ between $\mathrm{LG}_\mathit{00}$ and $\mathrm{LG}_\mathit{01}$ modes.  Please note that the estimated excitation probability refers to the particular LG mode.

After the writing process, a $100\,\mathrm{ns}$ long \textit{read} pulse having a $150\,\mathrm{\mu W}$ peak intensity was irradiated to the atomic ensemble.  Here the \textit{read} pulse was resonant on the atomic transition.  In all of the experiments the variable delay time $\delta t$ was set to $100\,\mathrm{ns}$.  Horizontally polarized stokes photons were collected by SMF4 using an optical system similar to that used for anti-Stokes photons.

Two single mode fibers (SMF1, SMF3) were aligned so that the \textit{write} and the \textit{read} pulses counter-propagating along the $z$ axis shared the same single spatial mode. Similarly, anti-Stokes and Stokes photons counter-propagating along the $z'$ axis shared a single spatial mode when the corresponding mode conversions were chosen for the holograms. The crossing angle between the $z$ and $z'$ axes was set to about $2.5^\circ$ in order to spatially separate the weak anti-Stokes (Stokes) photons from the strong \textit{write} (\textit{read}) pulses \cite{Harris2005}. 

The anti-Stokes and Stokes photons coupled into SMF2 and SMF4 were directed onto silicon avalanche photodiodes D1 and D2 (Perkin-Elmer model SPCM-AQR-14; detection efficiency, $0.62$) for photon counting. The outputs of D1 and D2 were fed to the start and the stop inputs of the time interval analyzer (TIA).
Fig.2-(a) shows the experimental results for the time-resolved coincidence over $1000\,\mathrm{s}$ between the anti-Stokes photons and Stokes photons of $\mathrm{LG}_\mathit{00}$ mode, from which the normalized cross intensity correlation function was estimated to be $g_\mathit{AS,S}(0)=22.2\pm 2.9$.  The repetition rate of our experiment was $4.5\times 10^5\,\mathrm{s}^{-1}$.  The count rate for the anti-Stokes photons was $3.1\times 10^2\,\mathrm{s}^{-1}$ and the coincidence count rate was $2.0\,\mathrm{s}^{-1}$. The measured transmission efficiency from the atomic ensemble to the detector was about $0.17$.  The estimated excitation probability with respect to the particular LG mode is  about $6.6\times 10^{-3}$, which confirms that the detected photons were well approximated by the single photon state.   

\begin{figure}[h]
  \begin{center}
    \includegraphics{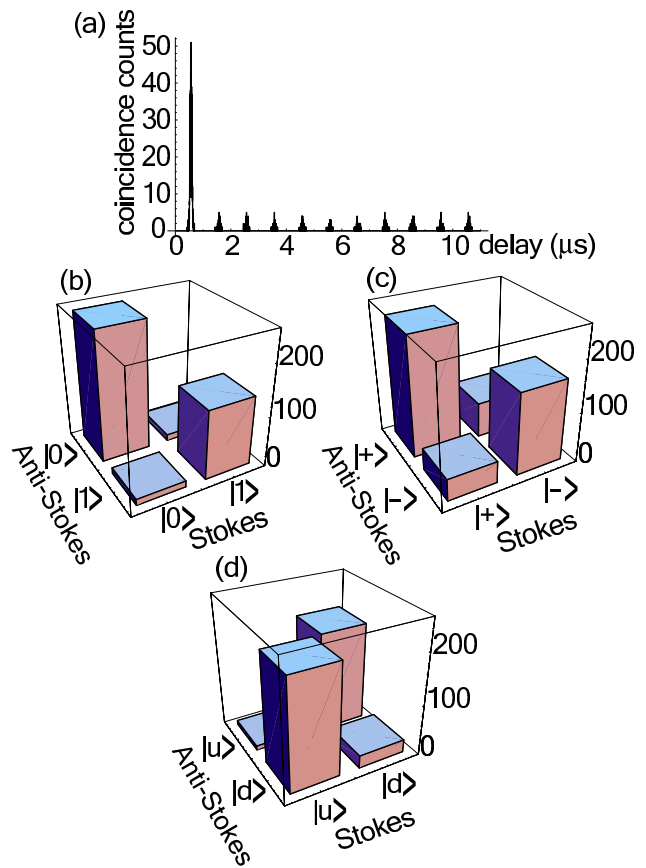}
  \end{center}
  \caption{(color online).  (a) Time-resolved coincidence counts, where the time resolution was set to $1.6\,\mathrm{ns}$.  (b), (c) and (d) are the coincidences in the various measurement basis.}
    \label{fig:2.eps}
\end{figure}

\section{ENTANGLEMENT ESTIMATION}
In order to evaluate the entanglement of OAM states, we investigated the coincidence counts under the various measurement bases. Here we utilize the following notation: $\ket{+}\equiv(\ket{0}+\ket{1})/\sqrt{2}$, $\ket{-}\equiv(\ket{0}-\ket{1})/\sqrt{2}$, $\ket{u}\equiv(\ket{0}+\mathrm{i}\,\ket{1})/\sqrt{2}$ and $\ket{d}\equiv(\ket{0}-\mathrm{i}\,\ket{1})/\sqrt{2}$.  Figs.2-(b), (c), and (d) graphically show the coincidence counts between anti-Stokes photons and Stokes photons, where the measurement was performed for four possible combinations of the three sets of orthogonal states $\{\ket{0},\ \ket{1}\}$, $\{\ket{+},\ \ket{-}\}$, and $\{\ket{u},\ \ket{d}\}$.  In the experiment, drifting of the optical alignment was inevitable due to the complexity of the system and each measurement was thus performed over a relatively short duration ($100\,\mathrm{s}$).  Fig.2-(b) clearly shows correlation of OAM between anti-Stokes and Stokes photons arising from angular momentum conservation law.  Note that such a correlation can be obtained even in the case of a simple mixture of $\ket{0}_\mathrm{AS}\ket{0}_\mathrm{S}$ and $\ket{1}_\mathrm{AS}\ket{1}_\mathrm{S}$ states.  The characteristics of entanglement can be found from Figs.2-(c) and (d), where strong correlation was observed on the basis with superposition states.  In order to verify whether the system was really entangled or not, we calculated a lower bound of the fidelity to the maximally entangled state $(\ket{0}_\mathrm{AS}\ket{0}_\mathrm{S}+\ket{1}_\mathrm{AS}\ket{1}_\mathrm{S})/\sqrt{2}$ from the coincidences determined from Fig.2-(c) and (d) \cite{imoto2002}. The lower bound is given by $p+q-1$, where $p$ \{$q$\} is the diagonal fraction in Fig.2-(c)\{(d)\}, and is calculated as $0.70\pm 0.08>0.5$.  This confirms entanglement between anti-Stokes and Stokes photons. From these results, entanglement of the OAM states between anti-Stokes photon and the atomic ensemble was verified.

\begin{figure}[h]
  \begin{center}
    \includegraphics{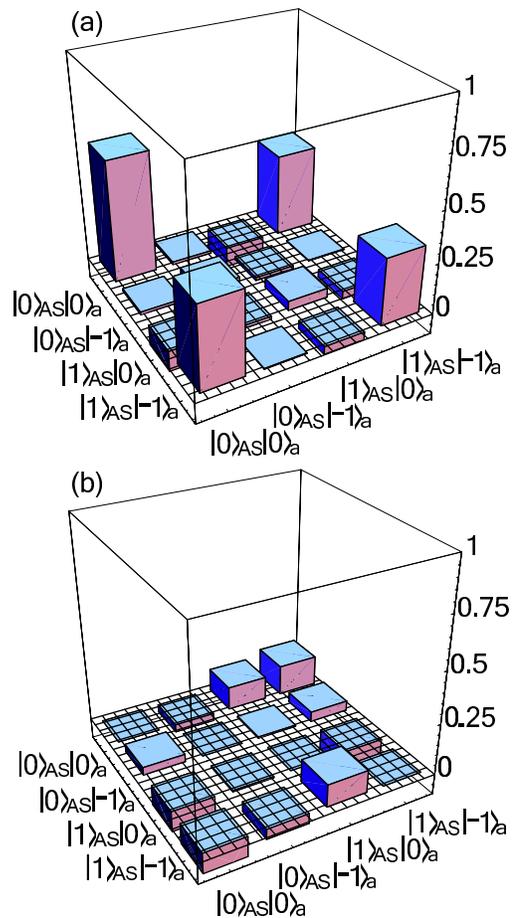}
  \end{center}
  \caption{(color online).  Graphical representation of the reconstructed density matrix.  (a) is the real part and (b) is the imaginary part. }
  \label{fig:3.eps}
\end{figure}
For the determination of the full state of the atoms and the anti-Stokes photon, we performed two-qubit state tomography \cite{Usami2003,White2001}, where the density matrix was reconstructed from the experimentally obtained coincidences for various combinations of the measurement basis. A graphical representation of the reconstructed density matrix is shown in Fig.3.  From the density matrix, the entanglement of formation (EOF) and the purity are estimated to be $0.76\pm 0.17$ and $0.92$, respectively.

Here we discuss the main sources which caused the errors for the lower bound of fidelity to the maximally entangled state and the entanglement of formation.  When the dislocation of the hologram is on the optical axis and the index of orbital angular momentum of a photon is changed, distribution of radial components also changes.  In contrast, our calculation says, when the dislocation is shifted from the optical axis by some amount and one to one superposition of $0-$ and $\hbar-$OAM states is generated through diffraction, distribution of radial components is almost identical to that for the incident photon.  When we estimated the lower bound of fidelity, we used only the one to one superposition measurement bases of $0-$ and $\hbar-$OAM states and we could thus neglect the effect of variation on the distribution of radial components.  Dominant factor of error simply came from Poisson fluctuation in the number of coincidence counts.  On the other hand, since we could not perform the two-qubit tomography simply by using such superposition measurement bases, the error of EOF came from both Poisson fluctuation in the number of coincidence counts and the variation on the distribution of radial components.  For example, the diagonal elements of Fig.2-B do not exactly correspond to the relative probability between $\ket{0}_\mathrm{AS}\ket{0}_\mathrm{S}$ and $\ket{1}_\mathrm{AS}\ket{1}_\mathrm{S}$, because a single-mode fiber only selects out the $p=0$ part of the radial components. Fortunately, a theoretical estimate shows that the difference in the radial components of OAM causes relatively small errors (it is less than $10{\%}$) which is comparable to the error originated from Poisson fluctuation in the number of coincidence counts. In other words, the obtained coincidences can be approximated to give the relative probability between $\ket{0}_\mathrm{AS}\ket{0}_\mathrm{S}$ and $\ket{1}_\mathrm{AS}\ket{1}_\mathrm{S}$.

The EOF obtained was relatively small compared with typical values for the pair of photons generated by the PDC process \cite{White2001}.  
The fidelity between pure state $\ket{\Psi(\gamma)}$ defined as Eq.(\ref{expected}) and reconstructed mixed state $\rho$, which is defined as $\bra{\Psi(\gamma)}\rho\ket{\Psi(\gamma)}$, is maximized when $\gamma=0.74\mathrm{e}^{\mathrm{i}0.11\pi}$, i.e., $\ket{\Psi(\gamma_{0})}\equiv\ket{\Psi(\gamma=0.74\mathrm{e}^{\mathrm{i}0.11\pi})}$ corresponds to the pure state approximation of the reconstructed density matrix.  Since EOF estimated from $\ket{\Psi(\gamma_{0})}\!\!\bra{\Psi(\gamma_{0})}$ is $0.94$ which is much larger than that for the reconstructed mixed state ($0.76$), the dominant factor limiting EOF is not the deference in the relative amplitude but the lowness of the purity.  

An atomic coherence time between two ground states which was not sufficiently longer than the storage time (delay time  $\delta t=100\,\mathrm{ns}$) may have caused the quantum correlation to decay and decreased the purity. The transit time of atoms passing through the beam area was estimated to be about $1\,\mathrm{ms}$ by assuming the Doppler limited temperature ($\sim 140\,\mathrm{\mu K}$)) and this value is apparently much longer than the delay time. Lamor precession cycle of the ground state Zeeman sublevels caused by the residual magnetic field is roughly estimated to be the order of $5\,\mathrm{\mu s}$, which will be the main origin of reduction in the purity.  By suppressing the eddy current, we would be able to prolong the coherence time and increase EOF as well as purity. 

While the difference about the relative amplitude between $\ket{0}_\mathrm{AS}\ket{0}_\mathrm{a}$ and $\ket{+1}_\mathrm{AS}\ket{-1}_\mathrm{a}$ states did not effect the EOF so much in two dimensional case, it should be balanced to demonstrate higher-dimensional entanglement. Such a balancing will be achieved by optimizing the focusing of laser beams \cite{Torner2003}.  In our scheme, an atom-photon entanglement is converted to a photon-photon entanglement.  Therefore, recently developed technique to check higher dimensional entanglement for twin photons \cite{SSR2004,SSR2005,SSR2006} could be applied to verify higher dimensional atom-photon entanglement.

\section{CONCLUSION}
In summary, we demonstrated entanglement of OAM states between a collective atomic excitation and a photon.  Storage of a quantum state involving the spatial mode associated with electromagnetically induced transparency \cite{Kuzmich2005-2} will enable us to share the higher-dimensional entanglement between two remote atoms.  Since, in principle, the quantum mechanical phase of atoms can be manipulated by applying a far off resonant laser beam, entanglement concentration \cite{AZ2003-2} in an atomic ensemble could also be achieved.  The experiment described in this paper raises the possibility that the atomic ensemble could be projected to any desired OAM state through the measurement of an anti-Stokes photon.  By using a Bose-Einstein condensate as an atomic ensemble, any desired yrast states \cite{Mottelson1999} could be generated in the range of  $-N\hbar$ to $N\hbar$.  

\section{ACKNOWLEDGEMENT}
We gratefully acknowledge M. Ueda, K. Usami, K. Akiba, A. Wada and T. Kishimoto for their valuable comments and stimulating discussions.
We also thank T. Yonemura, K. Bito and M. Moriya for supplying computer generated holograms.
This work was supported by Grant-in-Aid for Young Scientists (A), and the 21st Century COE Program at Tokyo Tech ``Nanometer-Scale Quantum Physics" by the MEXT.

\end{document}